# Cascading Neural Network Methodology for Artificial Intelligence-Assisted Radiographic Detection and Classification of Lead-Less Implanted Electronic Devices within the Chest

Mutlu Demirer, PhD, MBA[1]

Richard D. White, MD, MS[1]

Vikash Gupta, PhD[1]

Ronnie A. Sebro, MD, PhD[1]

Barbaros Selnur Erdal, DDS, MS, PhD[1]

[1]Center for Augmented Intelligence in Imaging-Department of Radiology, Mayo Clinic, Jacksonville, FL

Corresponding Author:

Barbaros Selnur Erdal, DDS, MS, PhD
Technical Director - Center for Augmented Intelligence in Imaging
Department of Radiology, Mayo Clinic
4500 San Pablo Road, Jacksonville FL 32224
Office: 904-953-6618
E-Mail: erdal.barbaros@mayo.edu

**ABSTRACT**

**Background & Purpose:** Chest X-Ray (CXR) use in pre-MRI safety screening for Lead-Less Implanted Electronic Devices (LLIEDs), easily overlooked or misidentified on a frontal view (often only acquired), is common. Although most LLIED types are "MRI conditional": 1. Some are stringently conditional; 2. Different conditional types have specific patient- or device- management requirements; and 3. Particular types are "MRI unsafe". This work focused on developing CXR interpretation-assisting Artificial Intelligence (AI) methodology with: 1. 100% *detection* for LLIED presence/location; and 2. High *classification* in LLIED typing.

**Materials & Methods:** Data-mining (03/1993-02/2021) produced an AI Model Development Population (1,100 patients/4,871 images) creating 4,924 LLIED Region-Of-Interests (ROIs) (with image-quality grading) used in Training, Validation, and Testing. For developing the cascading neural network (*detection* via Faster R-CNN and *classification* via Inception V3), "ground-truth" CXR annotation (ROI labeling per LLIED), as well as inference display (as Generated Bounding Boxes (GBBs)), relied on a GPU-based graphical user interface.

**Results:** To achieve 100% LLIED *detection*, probability threshold reduction to 0.00002 was required by Model 1, resulting in increasing GBBs per LLIED-related ROI. Targeting LLIED-type *classification* following *detection* of all LLIEDs, Model 2 multi-classified to reach high-performance while decreasing falsely positive GBBs. Despite 24% suboptimal ROI image quality, *classification* was correct in 98.9% and AUCs for the 9 LLIED-types were 1.00 for 8 and 0.92 for 1. For all mis*classification* cases: 1. None involved stringently conditional or unsafe LLIEDs; and 2. Most were attributable to suboptimal images.

**Conclusion:** This project successfully developed a LLIED-related AI methodology supporting: 1. 100% *detection*; and 2*.* Typically 100% type *classification*.

**INTRODUCTION**

Intra-thoracic placement of Lead-Less Implanted Electronic Devices (LLIEDs) capable of 1. Cardiac pacing, 2. Electrocardiographic recording, 3. Cardiovascular physiologic surveillance, or 4. Non-cardiovascular chemical monitoring has become commonplace.[1,2] The awareness of the presence of an implanted LLIED from the standpoint of both its general category (e.g., pacing *vs.* recording) and its specific type is critical to patient safety, LLIED function, clinical support operations, and/or local environmental hazards. This need for LLIED recognition is especially pertinent to the increasingly common electromagnetic and radiofrequency exposures during clinical Magnetic Resonance Imaging (MRI) examinations.[3]

Although most LLIEDs are considered to be "MRI conditional" (by posing no hazards in a specified MRI environment within specified conditions of use),[4] it remains imperative to acknowledge key facts about LLIEDs. These include the following realities: 1. MRI conditional does not mean MRI compatible or safe;[5] 2. Not all MRI-conditional LLIEDs carry equivalent potential risks, partly related to the co-existence of other implants;[6] 3. Even when considered MRI conditional, MRI exposure may result in recordable patient-related effects from an implanted LLIED or cause detectable alterations in LLIED function;[7-10] 4. Some MRI-conditional LLIEDs are considered to be more stringently conditional than others;[11] and 5. Different MRI-conditional LLIED categories, including Lead-Less Pacemakers (LLPs) compared to Lead-Less Recorders (LLRs), typically convey specific requirements for patient and/or LLIED assessment or preparation.[4,12,13] Moreover, some LLIEDs are considered to be "MRI unsafe" (by posing a significant risk in all MRI environments).[11,14]

A Chest X-Ray (CXR) is a standard component of pre-MRI safety screening (for LLIEDs or other man-made objects in the chest).[15-21] Unfortunately, any LLIED could be overlooked on CXR due to their common small sizes (comparable to a AAA battery, but subject to projection-related distortions), especially in the presence of 1. Suboptimal radiographic technique; 2. Patient-related factors; or 3.

Obscuration by superimposed-external or abutting-internal metallic or electronic materials. In addition, LLIED categories or types might be confused with each other by the interpreting radiologist because of: 1. LLIEDs having remarkably similar appearances and positions on a frontal CXR (typically the only view acquired in emergency/trauma department or intensive care unit settings);[19,22] or 2. From lack of familiarity by a radiologist with LLIED-specific characteristics.[18,23,24] These fundamental issues are even more pertinent to the less publicized, much smaller, and more stringently MRI-conditional (e.g., Pulmonary Artery Pressure Monitor (PAPM) for heart failure)[15,19,25] and MRI-unsafe (e.g., Esophageal Reflux Capsule (ERC) for pH-monitoring)[2,14,26] LLIEDs, which can easily go unnoticed.

Consequently, this work focused on the development of AI methodology to assist the CXR-interpreting physician in prompt and correct LLIED *detection* and *classification* with the following goals: 1. 100% *detection* sensitivity for general LLIED presence and location; and 2. High *classification* accuracy in LLIED typing.

## MATERIALS AND METHODS

**Original Study Population:**

With prior Institutional Review Board approval (including waived patient consent), data-mining of our institution-wide data-storage and Electronic Medical Record (EMR) systems for patients (each with a unique Medical Record Number (MRN)) with any entries related to LLIED placement, evaluation, or discovery was conducted (spanning: 03/1993 - 02/2021). The produced LLIED type-specific lists of patients/MRNs were then expanded to include records of all associated CXR examinations (each with a unique Accession number (ACC)). This compiled diverse dataset reflected a total of 3,875 unique patients/MRNs from our institution, which is comprised of three geographically dispersed quarternary-referral institutions and a multi-state network of over 70 satellite hospitals or ambulatory clinics. These patients/MRNs were represented by a total of 21,532 separate digital CXR examinations/ACCs, each consisting of at least one frontal view in either a Postero-Anterior (P-A) or Antero-Posterior (A-P) projection. All resultant 44,349 digital CXR images (91% frontal and 9% lateral) had been obtained to support standard-of-care by means of direct-digital radiology or computed radiology technologies[27] using a range of generations of 18 different manufactures of fixed and/or portable CXR systems.

For each patient/MRN, all available digital images per CXR examination/ACC (whether or not an LLIED was found to be present) were downloaded from the institution-wide "deconstructed" Picture Archiving and Communication System (PACS) [Radiology Information System: Radiant from Epic (Verona, WI); Vendor Neutral Archive system: Synapse from TeraMedica/Fujifilm Medical Systems USA, Inc. (Wauwatosa, WI); and Viewer: Visage Imaging from Pro Medicus Ltd (Richmond, Australia)] to a secure shared drive supporting a local Graphical User Interface (GUI)[28] with project-specific modifications of the underlying commercial software [MeVisLab from MeVis Medical Solutions AG (Bremen, Germany)].

All CXR images were reviewed by an cardiothoracic radiologist with 36 years of experience (also serving as project "ground-truth" expert) for basic cataloging, including appropriate categorizing of all frontal ("P-A/A-P") and Lateral ("Lat") views. When a CXR examination/ACC was represented by more than one P-A/A-P image (also Lat images, when applicable), all images were tentatively included for further consideration. However, conditions for immediate elimination included: 1. Duplicates or manipulated secondary captures of any image; and 2. Lack of evidence of an LLIED in the case of a Lat view. The resulting large image dataset was otherwise "real-world", without exclusion of images due to suboptimal image quality by any definition.

**Image Annotation:**

For image annotation using the GUI,[28] the "ground-truth" expert referred to the data-mined lists (relying on database and EMR corroboration, as needed) to delineate the specific LLIED types within the LLP (2 types), LLR (5 types), PAPM (only 1 type), and ERC (only 1 type) categories represented in the previously described 3,875-patient/MRN extraction [Appendix 1]. Based on expert confirmation of the presence of the expected entity, each P-A/A-P view demonstrating an LLIED was correspondingly labeled using the highly interactive (positioning, sizing, labeling) color-coded Region-Of-Interest (ROI) capabilities of the GUI [Figure 1].

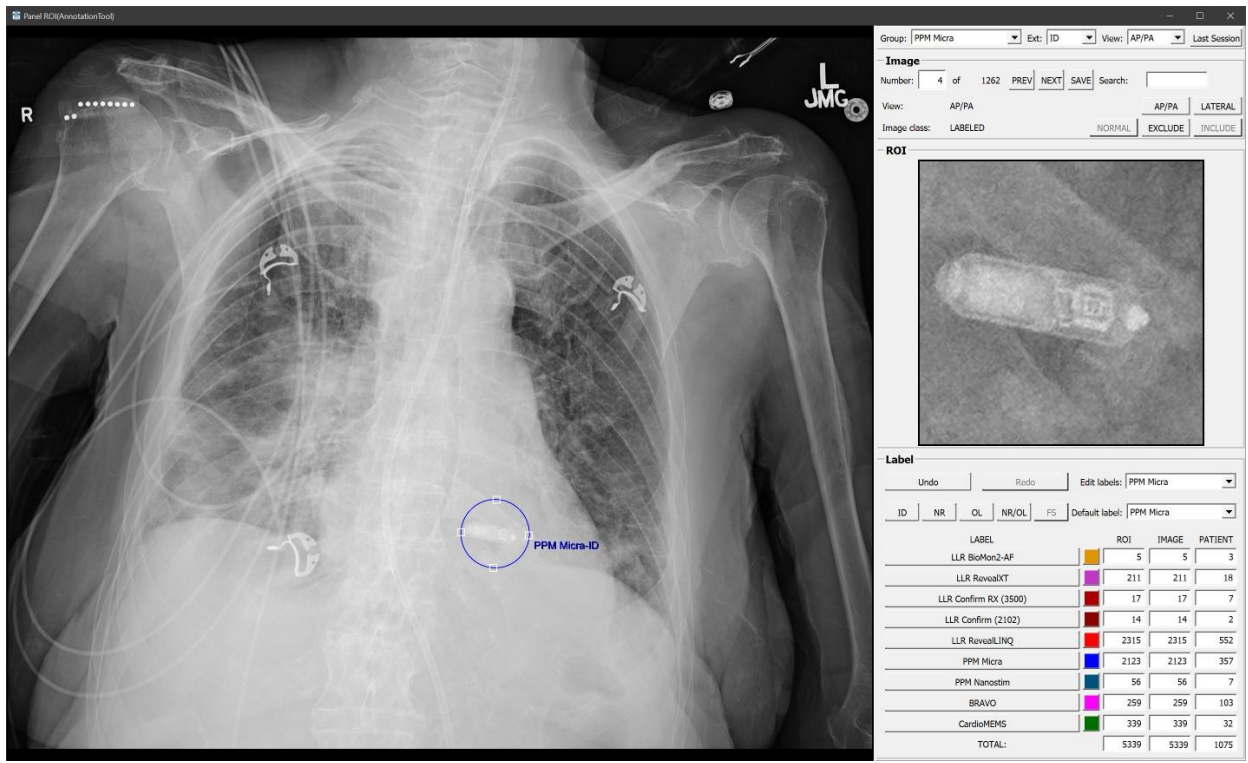

**Figure 1: GUI for CXR Annotation.**
The GUI allows the user to define the position and size of color-coded ROIs to indicate LLEID category and type, as well as assign an internal image-quality grade. The concurrent counts of annotated ROIs, images, and patients per LLEID label are shown, and user-directed filtering based on any of the aforementioned parameters supports focused reviews or revisions by the user.

In order to expand the data-subset sizes to help ensure quality in AI model development, the aforementioned provisionally acceptable Lat views were scrutinized by the "ground-truth" expert prior to LLIED labeling for possible inclusion only in model Training and Validation. If a Lat projection of a specific LLIED was considered (based on the absence of predetermined exclusion criteria [Appendix 2]) to be consistent with normal projectional variability often seen clinically on P-A/A-P views (e.g., due to differing patient and/or LLIED positioning), it was appropriately annotated for potential future use in a Training and Validation data subset [Appendix 3].

During the placement of ROIs to label one or more of the 9 LLIED types on any CXR image (P-A/A-P or acceptable Lat), a basic quality grade reflecting LLIED general conspicuity and detail clarity was applied per ROI as follows: 1. Unequivocally diagnostic with high device visibility and delineation, supporting reliable *detection* with localization and then *classification* for identification ("ID"); 2. Potentially Non-Recognizable ("NR") for *detection*, moreover for *classification*, due to poor device visibility (e.g., from suboptimal radiographic technique or motion-related blurring); 3. ID, but with externally superimposed or internally abutting radio-opaque man-made objects, or with incomplete P-A/A-P view inclusion within CXR-image margins, causing significant Over-Lapping ("OL") with obscuration of device boundaries or internal characteristics; or 4. Combined NR and OL ("NR&OL"). All ROIs, including those with suboptimal grades (i.e., graded NR, OL, or NR&OL), were included in AI model Training, Validation, and Testing processes.

**AI Model Development:**

**AI Technical Infrastructure:**

AI model development utilized several secure on-site Graphics Processing Unit (GPU) [Nvidia (Santa Clara, CA)]-dependent systems. Image-data curations and initial phases of model development relied on: 1. One workstation containing two GPUs [2 RTX 8000] with 96 GB total video memory, 128 GB system memory, 12 TBs of disk storage, and a 2 TB of SSD drive for OS Support (Windows 10); and 2. Two workstations containing a single GPU [RTX 8000] with 48 GB video memory, 128 GB system memory, 18 TBs of disk storage, and a 2TB SSD drive for Operating System Support (Windows 10).

For the final AI model Training, Validation, and Testing, a high-end eight-GPU system [DGX A100 from Nvidia (Santa Clara, CA)] was employed.

**AI Model Development Population:**

Ultimately, after the incorporation of only those patients/MRNs demonstrating CXR evidence of an LLIED at some point, 2,775 of the originally extracted 3,875 patients/MRNs were excluded. Thus, the resulting AI Model Development Population consisted of the remaining screened 1,100 unique patients/MRNs, represented by: 1. 3,553 examinations/ACCs, 2. 4,871 annotated digital CXR images, and 3. 4,924 LLIED ROIs (with ROI quality grades of: 3,763 (76%) ID; 579 (12%) NR; 472 (10%) OL; and 110 (2%) NR&OL).

| Table 1: AI Model Development Population Distribution | | | |
|---|---|---|---|
| | **Training** | **Validation** | **Testing** |
| Patients/MRNs (n=1,100) | 80% (n=877*) | | 20% (Total: n=223*) |
| Examinations/ACCs (n=3,553) | 75% (n=2,110**) | 25% (n=710**) | 100% (n=752) |
| CXR Images (n=4,871***) | 100% (n=3,006) | 100% (n=1,019) | 100% (n=874) |
| Original ROIs (n=4,924****) | 100% (n=3,027) | 100% (n=1,019) | 100% (n=878) |
| * Subsequent transfer of sufficient Patients/MRNs from Training/Validation subpopulation to Testing set, as needed, to achieve some representation of all LLIED types in Testing <br> ** Subsequent exchange of a few Examinations/ACCs for more optimal balance per labeled ROI <br> *** Exceeds the number of Examinations/ACCs due to multiple images per study <br> **** Exceeds the number of CXR Images due to multiple ROIs per image | | | |

For optimal use of the AI Model Development Population, the following standard approach to data distribution was used [Table 1][29]: 1. 80% of patients/MRNs were randomly selected to support only Training or Validation; the remaining 20% (with associated CXR examinations/ACCs) were preserved and, thus, avoided any patient-memory during their use which was restricted to Testing; and 2. Within the Training/Validation sub-population of patients/MRNs, associated CXR examinations/ACCs (considered distinctly different events due to variations in: date/time or site of examination, patient clinical status, patient position or inspiration, and/or radiographic system or parameters) were pooled before being randomly distributed to form the: 1. Training dataset, consisting of 75% of unique examinations/ACCs; and 2. Validation dataset, containing the remaining 25% of unique examinations/ACCs. Using this format, ROIs were distributed per specific label for LLIED type. However, in the infrequent instance of insufficient numbers of different patients/MRNs per LLIED type (often legacy LLIEDs) to support the patient-based 80:20 split, an exception was permitted. In such cases, the available few patients/MRNs could be represented in either Training (and possibly also Validation) or Testing, as long as there were sufficient numbers of distinctly different ACCs (e.g., 1 patient/MRN with 2 examinations/ACCs used only in Training and a different patient/MRN with 1 examinations/ACC used only in Testing) [Table 1]. When possible, any individual Training and Validation dataset-size imitations were alleviated by utilizing Lat-view ROIs as previously described [Appendix 3].[29] In addition, dataset-size imbalances were corrected through additional expansions using unique ROI variants generated by traditional image-augmentation techniques (consisting of permutations of ROI vertical flipping, horizontal flipping, width shifting (±20%), height shifting (±20%), channel shifting (±20%), shearing (±20%), zooming (±20%) and rotation (±20 degrees)) [Appendix 3].[42]

Hence, we were able to classify even LLIED types with limited sample sizes (often legacy LLIEDs) during model development.

**AI Model Training, Validation, and Testing:**

A 2-Tiered approach to AI model development was used: 1. First, to emphasize the *detection* of the general presence and location of any LLIED, and then 2. Second, to support *classification* of the specific type of LLIED represented if an LLIED had been detected. Ultimately, this led to the development of our cascading neural network methodology. For both Tiers of this methodology, all model preparation was performed using Keras-2.1.4[30] with TensorFlow-1.15.[31] The initial learning rate was 0.001 on a stochastic gradient descent optimizer[32] with a batch size of 16; re-training was terminated after 100 epochs. During the Training and Validation process, model performance (monitoring binary cross-entropy) on the Validation set was observed per epoch with preservation of the model of highest performance accuracy to that point; if the Validation accuracy increased in subsequent epochs, the model was updated.

Tier 1: LLIED *Detection*:

For the *detection* with localization of any LLIED-related ROIs, a Region-based Convolutional Neural Network (R-CNN) was used. To that end, Faster R-CNN ResNet-50 FPN[33,34] was specifically selected as the base algorithm, pre-trained on the MS COCO[35] 2017 dataset, and fine-tuned using 1-class (i.e., all LLIEDs together forming a single class) Training and Validation datasets [Appendix 3]. Inherent to this method was the output of inference results as Generated Bounding Boxes (GBBs).

Promoting a prerequisite mandate to detect all LLIEDs and miss none, probability threshold reductions from the standard 0.50 level were made, as needed, to overcome suboptimal image quality and achieve the desired 100% *detection* sensitivity in the Validation dataset prior to Testing. The consequential disadvantage was the expected excessive production of GBBs with high numbers of falsely positive inference results and the resulting poor positive prediction. However, to avoid in advance the likely inference output of extremely large and/or highly asymmetrical GBBs compromising localization, a GBB size/shape output filter was applied; this filter restricted output-GBB size to 15–120 mm in either dimension with an aspect ratio of 0.7–1.4 [Figure 2]. In addition, Non-Maximum Suppression was applied to suppress overlapping of GBBs with Intersection over Union (IoU)[36] greater than 0.4 [Figure 2].

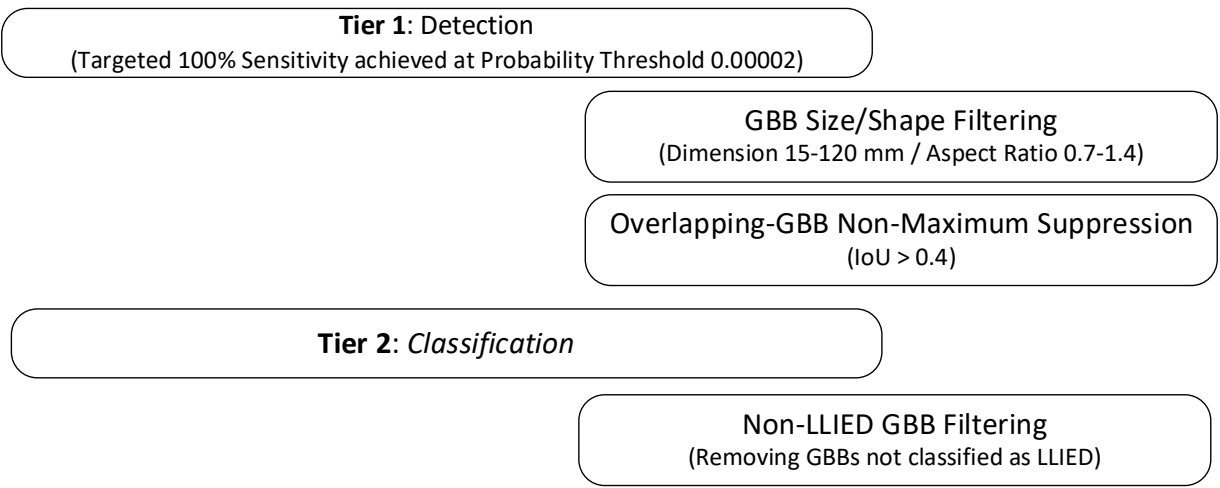

**Figure 2: Flow of Detection and Classification Processes**

For LLIED *detection* a: 1. True Positive (TP) inference result was recorded when a GBB overlapped with a ground-truth LLIED-related ROI at IoU $\geq$ 0.5; 2. False Positive (FP) resulted from a GBB not overlapping at IoU $\geq$ 0.5; and 3. False Negative (FN) resulted from the failure to create a GBB. Based on insights from the *detection* sensitivity evaluation of the Validation set in the first model (i.e., based on Faster R-CNN ResNet-50 FPN), the probability threshold was reduced from the standard level of 0.50 to 0.00002 prior to Testing to achieve the targeted *detection* sensitivity of 100% (i.e., recall value = 1.00).

Tier 2: LLIED-Type *classification*

With the combined goals of: 1. Reducing the number of false-positive results from Tier 1; and 2. Supporting maximal *classification* of the specific LLIED types, all LLIED *detection*-related GBBs were then classified using a multi-class CNN based on Inception V3.[37] Following transfer learning of initial weights derived from the ImageNet dataset to the base CNN,[38] its final layers were replaced by a fully connected layer of 1024 nodes in a ReLU activation unit,[39] followed by sigmoid output functions for multi-class classification.[40,41] The model weights were fine-tuned using ground-truth ROIs for the 9-class classifier (per specific LLIED type).

For per-case assignment of the LLIED-type, as determined by the classifier, the inference label on the GBB demonstrating the greatest IoU ($\geq$ 0.5) with the ground-truth LLIED-related ROI was used.[36]

**Statistical Analysis:**

As part of standard analysis of Testing results related to general LLIED *detection* in Tier 1, Precision-Recall Curves were plotted to reflect the basic comparison between the AI model output and "ground-truth" expert determinations [Figure 3].[43, 44]

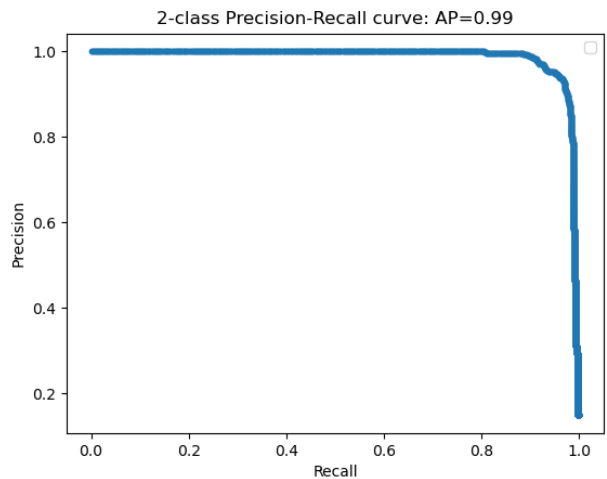

**Figure 3: Precision-Recall Curve for 2-Class Detection Model in Tier 1**.
Average Precision (AP) is indicated.

Tier-2 assessment of the discrimination performance of the AI model for multi-classifier *classification* used the Area Under the Receiver Operating Characteristic Curve (AUC ROC) methodology.[45-47]

**RESULTS**

**Model LLIED *Detection*:**

Along with the achievement of 100% LLIED *detection* sensitivity in the Testing set [Table 2], the required decreased probability threshold of 0.00002 for the first model resulted in increasing numbers of GBBs (including overlapping TPs or non-overlapping FPs) per LLIED-related ROI (average GBB per detected-LLIED ROI: 950/853 = 1.1 at threshold 0.50, increasing to 5,359/878 = 6.1 at threshold 0.00002).

| Table 2: Tier-1 Testing Results - LLIED Detection |||||||
|---|---|---|---|---|---|---|
| **LLIED-Related ROIs** | **Probability Threshold** | **GBBs** ||| **ROIs Undetected (FN)** | **Detection Precision (TP/(TP+FP))** | **Detection Sensitivity (TP/(TP+FN))** |
| | | **Total** | **TP** | **FP** | | | |
| 878 | 0.50 | 950 | 853 | 97 | 25 | 0.90 | 0.97 |
| | 0.00002 * | 5,359 | 878 | 4,481 | 0 | 0.16 | 1.00 |
| Probability Threshold: Probability threshold for LLIED-related ROI detection  
GBB: Generated Bounding Box  
TP: True Positive inference result when a GBB overlapped an LLIED-related ROI at IoU $\geq$ 0.5  
FP: False Positive inference result when a GBB did not overlap an LLIED-related ROI at IoU $\geq$ 0.5  
FN: False Negative result from the failure to create a GBB for an LLIED-related ROI  
* Probability threshold applied for GBB in Tier-1 Testing |||||||

**Model LLIED-Type *Classification*:**

With the goal of achieving maximal LLIED-type *classification* accuracy following the initial mandatory *detection* of all LLIEDs, the second model was used as a multi-classifier. After classifying the 5,359-GBB output from Tier-1 Testing, the number of FP GBBs was decreased by 3,462 GBBs (via initial *classification* as non-LLIEDs) while the mandated 100% *detection* sensitivity was preserved [Table 3]. Thus, no LLIED-related ROIs were missed and, of those classified as LLIED types, the *classification* assignments were correct in 868 of 878 or 98.9% of LLIED-related ROIs.

| Table 3: Tier-2 Testing Results - LLIED-Type Classification |||||||
|---|---|---|---|---|---|---|
| **LLIED-Related ROIs** | **GBB Classification** |||| **Detection Precision** | **Detection Sensitivity** ||
| | **Total Tier-1 Output** 5,359 |||| | ||
| | **LLIED** ||| **Non-LLIED** | | ||
| 878 | 1,897 ||| 3,462 | 0.46 | 1.00 ||
| | **Correct Type** | **Incorrect Type** | **Confirmed Non-LLIED** | 4,481 | | |
| | 868 (98.9%) | 10 (1.1%) | 1,019 | | | |

AUCs for the *classification* of the LLIED-type were 0.92 for 1 type (an MRI conditional LLR type) and 1.00 for 8 types (including stringently MRI-conditional PAPM and MRI-unsafe ERC types) (Figure 4). When the 10 detected but misidentified cases (10/878 = 1.1% of all LLIED-related ROIs) were considered further, the following characteristics were noted: 1. None involved the mis*classification* of either an MRI-stringently conditional PAPM or MRI-unsafe ERC LLIED; and 2. Eight cases of mis*classification* of an LLIED-related ROI could be attributed to suboptimal image-quality grades (4 NR&OL, 2 NR, and 2 OL).

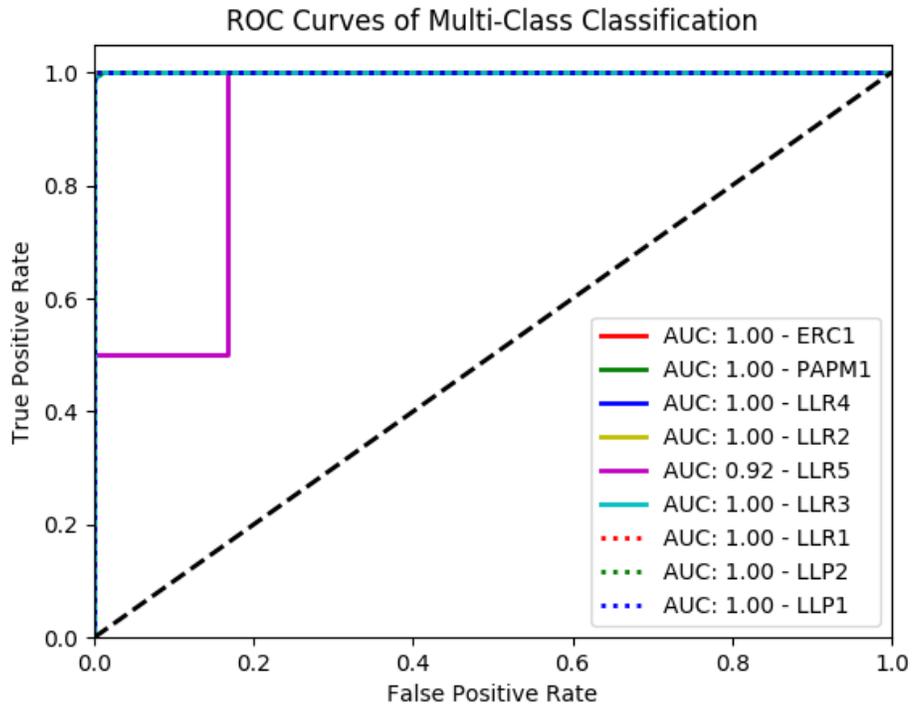

**Figure 4: ROC Curve for Multi-Class Classification Model in Tier 2**.

## DISCUSSION

In our study, we focused on developing AI methodology to potentially assist the physician interpreting a digital frontal CXR in the *detection* with localization, as well as *classification*, of a range of commonly implanted LLIEDs to support related pre-MRI safety screening. To our knowledge, this is the first reported achievement of an AI-based radiographic *detection* and *classification* system directed at the array of LLIEDs, ranging from MRI-conditional to MRI-unsafe, that may be found on CXR (at times incidentally) to be implanted in patients.

Our developed cascading neural network methodology for AI model development first achieved in Tier 1 the pre-determined mandated *detection* and location of any LLIED with 100% sensitivity in the AI Model Development Population evaluation. To that end, a Faster R-CNN approach[33,34] was selected over other popular detection/localization approaches (e.g., YOLO[48]) due to the relative adequacy of its speed but its superior accuracy.[49] The reduction in probability threshold required to detect all LLIEDs produced multiple FP GBBs which were well compensated for by combined size/shape-based filtering and the strength of the multi-classifier in Tier 2 processing.

Next, in Tier 2, the methodology utilized an InceptionV3-based multi-class CNN as a multi-classifier to achieve very high *classification* accuracies of known LLIEDs (i.e., previously classified within a model) in the AI Model Development Population evaluation (i.e., AUCs = 0.92 for 1 type and 1.00 for 8 types). No cases of LLIED-type mis*classification* involved either an MRI-stringently conditional or MRI-unsafe type, and most mis*classification* cases could be attributed to suboptimal image quality.

The importance of continuous learning for AI-model improvement[50] was reinforced in this project. It is also crucial to recognize that this project was consistent with a real-world experience[50] by its: 1. Utilization of a large dataset representing multiple geographically dispersed sites; 2. The presence of all levels of general radiographic quality from multiple systems producing digital CXRs over almost 3 decades; and 3. Inclusion of all levels of LLIED image quality (with NR, OL, and NR&OL cumulatively accounting for 24% of the LLIED representations in the AI Model Development Population).

Limitations:

There are limitations to our study. First, there is currently a need to execute our 2 cascading models at very low probability thresholds to prevent potentially failing to detect all LLIEDs due to suboptimal overall CXR image quality. This leads to the creation of additional GBBs potentially suggesting the presence of an LLIED in a nonLLIED case. A potential future consideration would be to adjust model parameters based on overall CXR image quality on a case-by-case basis. For example, if the overall signal-to-noise ratio is poor, the cascading models could be executed with lower probability thresholds; otherwise, thresholds could remain at a traditional 0.50. However, additional probability threshold adjustments might still be warranted even when the overall signal-to-noise ratio is good, but LLIEDs are obscured by adjacent or superimposed prominent soft tissues (e.g., in the lower chest or upper abdomen). Nonetheless, with the achievement and maintenance of 100% *detection* sensitivity, a fundamental priority in this project, such labor-intensive optimization of the user experience by further reduction in FP GBB display was left to future refinements during clinical deployment.

Second, while this work represents a single-institution experience with inherent potential population data bias (while the LLIEDs have set designs), it is important to recognize that the institution is comprised of many geographically dispersed clinical sites (approximately 75) which contributed over many years (almost 30) via a common IT infrastructure to create our large pooled Original Study Population (almost 4,000 patients, from which 1,100 within the AI Model Development Population had CXR examinations demonstrating LLIEDs). Despite the aforementioned population attributes controlling data bias, the resulting AI Model Development Population represented an abnormally high background prevalence of LLIEDs, potentially impacting positively on reported AI model performance.[47]

**Conclusion:**

This project successfully led to the development of AI methodology reaching important goals, including: 1. 100% *detection* sensitivity for general LLIED presence and location, and 2. High *classification* accuracy in LLIED typing and, by default, MRI-safety level determination.

# APPENDICES

| LLIED Categories and Types Represented in Original Study Population |||||
|---|---|---|---|---|
| **LLIED** || *EMA*/FDA Approval | **MRI Safety**[A-D] ||
| **Category** | **Type** | | **1.5 Tesla** | **3.0 Tesla** |
| **LLP** | 1[A] | *10/2013* | Conditional | Conditional |
|  | 2[B] | 04/2016 | Conditional | Conditional |
| **LLR** | 1[B] | 11/2007 | Conditional | Conditional |
|  | 2[A] | 08/2008 | Conditional | *INA* |
|  | 3[B] | 02/2014 | Conditional | Conditional |
|  | 4[C] | 04/2016 | Conditional | Conditional |
|  | 5[A] | 09/2017 | Conditional | Conditional |
| **PAPM** | 1[A] | 10/2006 | Conditional* | Conditional* |
| **ERC** | 1[B] | 12/2010 | Unsafe | Unsafe |

LLIED = Lead-Less Implanted Electronic Device
EMA = European Medicines Agency
FDA = United States Food & Drug Agency
LLP = Lead-Less Pacemaker
LLR = Lead-Less Recorder
PAPM = Pulmonary Artery Pressure Monitor
ERC = Esophageal Reflux Capsule
[A] = http://www.mrisafety.com/List.html
[B] = https://www.abbott.com/for-healthcare-professionals.html
[C] = https://global.medtronic.com/xg-en/healthcare-professionals.html
[D] = https://www.biotronik.com/en-gb/products/arrythmia-monitoring/biomonitor-2
Conditional = Safe if following specific recommendations or guidelines per manufacturer
Conditional* = Safe only if imaged under stringent and highly specific MRI technical restrictions
Unsafe = Unsafe in an MRI environment
*INA* = Information Not Available

**Appendix 1**
Categories and types (including MRI-safety levels) of LLIEDs
represented in Original Study Population, and ultimately in AI Model Development Population

| LLIED Projection-Related Exclusion Criteria on Lateral Views |||
|---|---|---|
| LLIED Type | Entity | Exclusion Criteria |
| LLP | 1 | Excessive foreshortening preventing:<br>• Simultaneous visualization of fixation helix and distal battery chevron[A,B] (and)<br>• Appearance of body length > 3 times diameter |
| LLP | 2 | Excessive foreshortening preventing:<br>• Simultaneous visualization of cathode/tine complex and electronics-battery transition zone (approximately 0.5 body length)[C] (and)<br>• Appearance of body length > 2 times diameter |
| LLR | 1 | Excessive foreshortening preventing:<br>• Simultaneous visualization of battery-electronics transition zone (approximately 0.4 body length) and electronics-antenna transition in rectangle-shaped body[D-F] (and)<br>Lack of en-face presentation facilitating:<br>• Visualization of rectangular distal electrode[D-F] |
| LLR | 2 | Excessive foreshortening preventing:<br>• Simultaneous visualization of battery-electronics transition zone (approximately 0.4 body length) and electronics-antenna transition in slightly teardrop-shaped body[D-E] (and)<br>Lack of en-face presentation facilitating:<br>• Visualization of triangular distal electrode[D-E] |
| LLR | 3 | Excessive foreshortening preventing:<br>• Simultaneous visualization of battery-electronics transition zone (approximately 0.3 distance) and electronics-antenna transition in rectangle-shaped body[F,G] (and)<br>Lack of en-face presentation facilitating either:<br>• Visualization of 3-dot pattern aligned along electronics board and antenna base[F,G]<br>(or) Visualization of corrugated-appearing medradio antennae supporting cellular communication[F,G] |
| LLR | 4 | Excessive foreshortening preventing:<br>• Simultaneous visualization of battery-electronic transition zone (approximately 0.4 body length) and faintly radio-opaque elongated antenna with distal electrode cap[H,I] (and)<br>Lack of en-face presentation facilitating:<br>• Visualization of 2 small projections from body at base of antenna[H,I] |
| LLR | 5 | Excessive foreshortening preventing:<br>• Simultaneous visualization of battery-electronics transition zone (approximately 0.5 body length) and electronics-antenna transition in rectangle-shaped body[H,J] (and)<br>Lack of en-face presentation facilitating either:<br>• Visualization of 2 projections to triangular antenna supporting bluetooth communication (or) Visualization of plaid-like pattern in battery[H,J] |
| PAPM | 1 | All included |
| ERC | 1 | All included |

A. J Cardiovasc Electrophysiol. 2016 Dec;27(12):1502-1504. doi: 10.1111/jce.13104. Epub 2016 Oct 26. PMID: 27704685.
B. Curr Cardiovasc Risk Rep 2018; 12, 11. https://doi.org/10.1007/s12170-018-0575-8
C. https://www.globalradiologycme.com/single-post/2019/03/25/micra-intracardiac-pacemaker
D. https://thoracickey.com/imaging-of-implantable-devices-2/
E. Curr Cardiol Rev. 2012 Nov;8(4):354-61. doi: 10.2174/157340312803760758. PMID: 22920479; PMCID: PMC3492818.
F. https://www.globalradiologycme.com/single-post/2015/11/03/implanted-cardiac-loop-recorder
G. https://fccid.io/LF5MEDSIMPLANT1/Operational-Description/Antenna-Info-2088509
H. J Arrhythm. 2018 Nov 20;35(1):25-32. doi: 10.1002/joa3.12142. PMID: 30805041; PMCID: PMC6373656.
I. Heart Lung Circ. 2018 Dec;27(12):1462-1466. doi: 10.1016/j.hlc.2017.09.005. Epub 2017 Oct 6. PMID: 29054505.
J. https://www.innovationsincrm.com/cardiac-rhythm-management/articles-2018/july/1273-trends-in-subcutaneous-cardiac-monitoring-technology

**Appendix 2**

Specific criteria for exclusion of an LLIED projection on lateral CXR view from AI model development

| | | | Available from P-A/A-Ps (A) | Plus Supplemental Lats (B) | Data Subset Subtotals (A+B) | Plus Augmented P-A/A-Ps (C) | Data Subset Totals (A+B+C) |
|---|---|---|---|---|---|---|---|
| **Labeled-ROI Dataset Expansion and Balancing** | | | | | | | |
| **LLIED** | **LLP** | **LLP1** Tr | 24 | 16 (of 16) | 40 | 1,186 | 1,226 |
| | | V | 7 | 7 (of 7) | 14 | 401 | 415 |
| | | T | 2 | | 2 | | 2 |
| | | **LLP2** Tr | 997 | 229 (of 329) | 1,226 | 0 | 1,226 |
| | | V | 317 | 88 (of 88) | 405 | 10 | 415 |
| | | T | 332 | | 332 | | 332 |
| | **LLR** | **LLR1** Tr | 82 | 46 (of 46) | 128 | 1,098 | 1,226 |
| | | V | 36 | 15 (of 15) | 51 | 364 | 415 |
| | | T | 35 | | 35 | | 35 |
| | | **LLR2** Tr | 3 | 3 (of 3) | 6 | 1,220 | 1,226 |
| | | V | 1 | 1 (of 1) | 2 | 413 | 415 |
| | | T | 5 | | 5 | | 5 |
| | | **LLR3** Tr | 1,226 | 0 (of 319) | 1,226 | 0 | 1,226 |
| | | V | 415 | 0 (of 92) | 415 | 0 | 415 |
| | | T | 441 | | 441 | | 441 |
| | | **LLR4** Tr | 1 | 1 (of 1) | 2 | 1,224 | 1,226 |
| | | V | 1 | 1 (of 1) | 2 | 413 | 415 |
| | | T | 1 | | 1 | | 1 |
| | | **LLR5** Tr | 9 | 2 (of 2) | 11 | 1,215 | 1,226 |
| | | V | 4 | 0 (of 0) | 4 | 411 | 415 |
| | | T | 2 | | 2 | | 2 |
| | **PAPM** | **PAPM1** Tr | 158 | 79 (of 79) | 237 | 989 | 1,226 |
| | | V | 49 | 18 (of 18) | 67 | 348 | 415 |
| | | T | 33 | | 33 | | 33 |
| | **ERC** | **ERC1** Tr | 101 | 50 (of 50) | 150 | 1,075 | 1,226 |
| | | V | 38 | 21 (of 21) | 59 | 356 | 415 |
| | | T | 27 | | 27 | | 27 |

Tr = Training / V = Validation / T = Testing
Tier 1 = Detection of LLIEDs
Tier 2 = Classification of specific LLIED types

**Appendix 3**
Dataset-size expansion and balancing for AI model development